# The Measurement of Near-Field Thermal Emission Spectra using an Infrared Waveguide

Saman Zare[*], Carl P. Tripp[**,†] and Sheila Edalatpour[*,†]

[*] *Department of Mechanical Engineering, University of Maine, Orono, ME 04469*
[**] *Department of Chemistry, University of Maine, Orono, ME 04469*
[†] *Frontier Institute for Research in Sensor Technologies, University of Maine, Orono, ME 04469*

**Abstract**

We describe a simple and robust method using an internal reflection element acting as an infrared waveguide to measure the spectra of near-field thermal emission. We experimentally demonstrate the spectrally-narrow peaks of near-field thermal emission by isotropic media due to the excitation of surface phonon-polaritons in quartz and amorphous silica and due to the frustrated total-internal-reflection modes in amorphous silica and polytetrafluoroethylene. Additionally, we demonstrate the broadband near-field thermal emission of hyperbolic modes in hexagonal boron nitride which is an anisotropic uniaxial medium. We also present a theoretical approach based on the fluctuational electrodynamics and dyadic Green's functions for one-dimensional layered media for accurate modeling of the measured spectra.

Many near-field applications, such as nano-gap thermophotovoltaic power generation [1-8] and thermal rectification [9-20], rely on spectrally-selective thermal emission in the near field. Most of the experimental studies on near-field thermal radiation have measured the total (spectrally integrated) heat transfer which does not provide information about the spectrum of heat transfer [7,21-47]. Narrow peaks can be observed in the near-field thermal emission by isotropic media due to the excitation of surface phonon- and plasmon-polaritons (SPhPs and SPPs, respectively) as well as frustrated total-internal-reflection modes. In addition to surface and frustrated modes, anisotropic media can support hyperbolic modes which enhance near-field thermal emission in a broadband manner. Measuring the spectrum of near-field thermal emission is challenging because the evanescent waves in the near-field need to be converted to propagating waves to reach a Fourier-transform infrared (FTIR) spectrometer located in the far zone. So far, the SPhP modes of near-field thermal emission have been measured for silica [48,49], quartz [50], silicon carbide (SiC) [48,50-52], and for a thin film of hexagonal boron nitride (hBN) on gold and silica substrates [52]. The frustrated total-internal-reflection modes thermally emitted by polytetrafluoroethylene (PTFE) have also been measured [50,52]. However, the broadband near-field thermal emission due to hyperbolic modes has not been experimentally observed.

Most of the measured near-field spectra are obtained using scanning optical microscopes [48,50-52]. In this spectroscopic technique, the thermal near field of an emitting sample is scattered to the far zone by bringing the sharp tip of the microscope to a sub-wavelength distance from the sample. The measured peaks using this technique are redshifted (from 3 cm$^{-1}$ to 63 cm$^{-1}$) and broadened relative to the theoretical predictions using the fluctuational electrodynamics [48-54]. It is shown experimentally [51], and theoretically [54], that the spectral redshift and broadening of the peaks are strongly dependent on the geometry of the probe. For example, three different values of 898

cm$^{-1}$, 923 cm$^{-1}$, and 943 cm$^{-1}$ were obtained for the SPhP resonance of SiC when different probes were used in the same experimental setup [51]. The reason for the three different values for the resonance arises from the difficulty in precisely controlling and measuring the geometry of the probe. Furthermore, computationally expensive numerical models are required for relating the measured signal to the near-field thermal emission by the sample.

In this study, we present a simple and robust spectroscopy technique which does not involve multiscale, complex-shape probes and specialized optical instruments. In addition to SPhP and frustrated modes for quartz, silica and PTFE, we experimentally demonstrate broadband hyperbolic thermal emission for hBN. The measured spectra are reproducible, and the redshift and broadening of the peaks can accurately be predicted using a theoretical model based on the fluctuational electrodynamics and dyadic Green's functions (DGFs) for one-dimensional layered media.

A schematic of the experimental setup for near-field thermal emission spectroscopy is shown in Fig. 1a. An internal reflection element (IRE) with a high refractive index and low infrared losses is brought into contact with an emitting sample. Due to surface roughness, several sub-wavelength air gaps are formed between the two surfaces. The air gap due to surface roughness is represented by *d* in Fig. 1b where a close-up view of the sample-IRE interface is shown (As it will be discussed later, the presence of an air gap is verified in Section VI of Supplemental Materials). The emitted evanescent waves with parallel component of the wavevector, $k_\rho$, between $k_0$ and $n_I k_0$ ($k_0$ and $n_I$ are vacuum wavevector and IRE refractive index, respectively) are converted into propagating waves in the IRE due to the increase of the wavevector (Fig. 1b). These propagating waves cannot escape the parallel surfaces of the IRE as they experience total internal reflections. Instead, these modes are guided through the IRE toward its beveled ends which make an angle of *α* with the surface normal. The coupled modes with $k_\rho$ between $k_0$ and $\sin(90° - \alpha + \theta_{cr})n_I k_0$, where $\theta_{cr} = \sin^{-1}(1/n_I)$ is the critical angle for the IRE-air interface, hit the IRE beveled ends with an angle smaller than the critical angle (see Section I of Supplemental Materials). These modes exit from IRE's beveled ends and are sent to an FTIR spectrometer where their spectrum is recorded. The higher the refractive index and the bevel angle of the IRE, the larger the number of evanescent modes that can couple to the IRE. Since the IRE is transparent in the infrared, its thermal emission is negligible compared to the sample. The waveguide arrangement shown in Fig. 1b is similar to the inverse of the Kretschmann configuration used for exciting SPPs at the interface of a metallic thin film and the free space using an external illumination [55].

The sample is mounted on a metal ceramic heater with an output of 24 W at 24 V (Thorlabs, HT24S) which is connected to a power supply (KEPCO, ABC 36-3DM) with a maximum voltage of 36 V. The sample temperature is measured using a K-type thermocouple and is read using a digital thermometer (OMEGA, HH-52). The sample-heater assembly is adhered to a ceramic base using a nickel-base metallic adhesive (Cotronics Corp., Durabond 952 FS). A zinc selenide (ZnSe) IRE with a trapezoidal cross section and a bevel angle of $\alpha = 45°$ is selected for the experiment (Harrick, EM2122). The IRE dimensions are 50 (length) × 10 (width) × 2 (thickness) mm. Zinc selenide has a refractive index of $n_I \approx 2.4$ and is transparent between 700 cm$^{-1}$ and 15000 cm$^{-1}$. The IRE is placed on the sampling surface of a multiple-reflection horizontal Attenuated Total Reflection (ATR) accessory (Harrick, HorizonTM). The base-heater-sample assembly is put in contact with the IRE while two elastic posts made of 0.005"-thick stainless-steel plates acting as a spring are placed between the base and the sampling plate of the ATR accessory. A pressure applicator is used to press the base toward the IRE for near-field measurements. For far-field measurements, the pressure is released from the base such that the elastic posts push the sample away and keep it at a 1-mm distance from the IRE. The signal exiting one of the IRE beveled ends is collected by the ATR accessory and transferred into the emission port of an FTIR spectrometer using a $f/4$ parabolic mirror. The FTIR is an ABB-Bomem MB1552E equipped with a broad band mercury-cadmium-telluride (MCT) detector (InfraRed Associates Inc.).

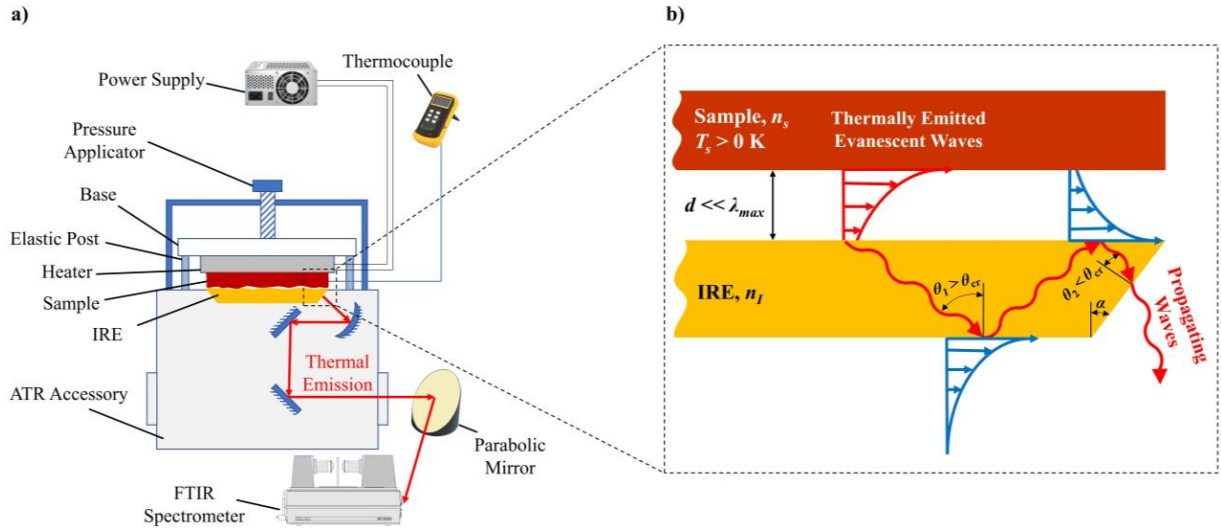

Figure 1 – (a) A schematic of the experimental setup for near-field thermal emission spectroscopy. (b) A close-up view of the sample-IRE interface. The distance $d$ represents the air gap between the sample and the IRE due to surface roughness. Thermally emitted evanescent waves with $k_\rho$ between $k_0$ and $\sin(90° - \alpha + \theta_{cr})n_I k_0$ exit the IRE's beveled ends after several total internal reflections, and are collected using an FTIR spectrometer.

Near-field and far-field thermal spectra are measured for quartz, silica, PTFE, and hBN at a temperature of approximately 160°C (see Section II of Supplemental Materials). The silica, hBN, and PTFE samples are 1-mm thick, while the thicknesses of the quartz sample is 0.5 mm. It is verified experimentally and theoretically that all samples are optically thick. A spectral resolution of 4 cm$^{-1}$ is selected for the FTIR spectrometer. The background thermal emission is measured by blocking the sample emission from reaching the FTIR via placing a thick film of stainless steel (which is opaque in the infrared) at the exit of the ATR accessory. The background thermal emission is subtracted from the measured signals. The near-filed spectra are normalized by the far-field thermal emission to compensate for the wavenumber-dependent responsivity of the photodetector, absorption by the internal parts of the ATR accessory and ambient gases, as well as modulation efficiency of the FTIR spectrometer (see Section III of Supplemental Materials). The normalized near-field spectra are shown in Fig. 2a to 2d. To ensure that the experiments are reproducible, the measurements were repeated three times for each sample. In each repetition, the IRE, the sample, and the heater are disassembled and reinstalled. As an example, the three measured spectra for the quartz sample are plotted in Fig. 2a. This level of repeatability was observed for all samples.

To compare measurements with theory, the energy density emitted by the samples in free space is calculated using the fluctuational electrodynamics and the DGFs for a semi-infinite bulk [56-58]. The near-field and far-field energy densities are calculated at 200 nm (approximately equal to the surface flatness of the samples) and 1 mm above the samples, respectively. The dielectric functions of quartz, silica, PTFE, and hBN are obtained from literature [59-61] and are plotted in Section IV of Supplemental Materials. It should be noted that hBN is an anisotropic uniaxial medium and is described using a parallel and a perpendicular (to the optical axis) dielectric function. The calculated near-field spectra normalized by their far-field value are plotted in Figs. 3e to 3h. Peaks in near-field thermal spectra occur when $\varepsilon_s + 1 \to 0$ (due to the excitation of SPhPs) and $\text{Im}[\varepsilon_s] \to \infty$ (due to the contribution of frustrated total-internal-reflection modes), where $\varepsilon_s = n_s^2$ is the dielectric function of the sample [62]. Additionally, thermal emission by uniaxial media can be enhanced in a broadband manner when $\text{Re}[\varepsilon_{s,\|}]\text{Re}[\varepsilon_{s,\perp}] < 0$ ($\varepsilon_{s,\|}$ and $\varepsilon_{s,\perp}$ are the dielectric functions of the sample in the parallel and perpendicular directions relative to the optical axis, respectively) due to the excitation of hyperbolic modes.

The near-field energy density for quartz (Fig. 3e) has three peaks at 806 cm$^{-1}$, 1155 cm$^{-1}$, and 1189 cm$^{-1}$ due to the excitation of surface phonon-polaritons (Re[$\varepsilon_s$] = -0. 5, -1.4, and -1.0, respectively). The three SPhP resonances are captured in the measured quartz spectrum (Fig. 3a). The SPhP resonances are redshifted and broadened in the experiments compared to the theoretical predictions. The origin of the redshift and broadening of the peaks will be discussed later using a theoretical model. The theoretical energy density for silica (Fig. 3f) shows two peaks at 809 cm$^{-1}$ and 1149 cm$^{-1}$ which both are measured (Fig. 3b). The low-wavenumber peak is due to the symmetric Si-O-Si stretching vibrations (frustrated modes), while the high-wavenumber peak is because of the excitation of SPhPs (Re[$\varepsilon_s$] = -1.2). The theoretical spectrum for PTFE (Figs. 3g) has two peaks at 1163 cm$^{-1}$ and 1225 cm$^{-1}$, which are due to the symmetric and asymmetric C-F stretching vibrations (frustrated modes). These peaks are also captured in the measured spectrum (Fig. 3c). The experimental and theoretical spectra of near-field thermal emission by hBN are shown in Figs. 3d and 3h, respectively. Hexagonal boron nitride has two hyperbolic bands at 780 cm$^{-1}$ – 835 cm$^{-1}$ and 1400 cm$^{-1}$ – 1600 cm$^{-1}$. These two hyperbolic bands are captured in the measured spectrum of hBN in Fig. 3d. The increase of the measured signal near 700 cm$^{-1}$ is due to thermal emission by ZnSe which becomes opaque around this wavenumber.

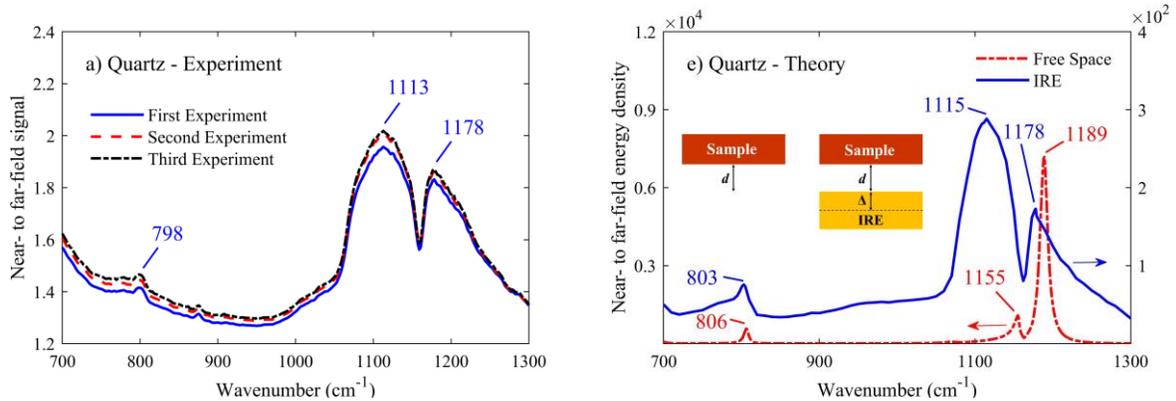

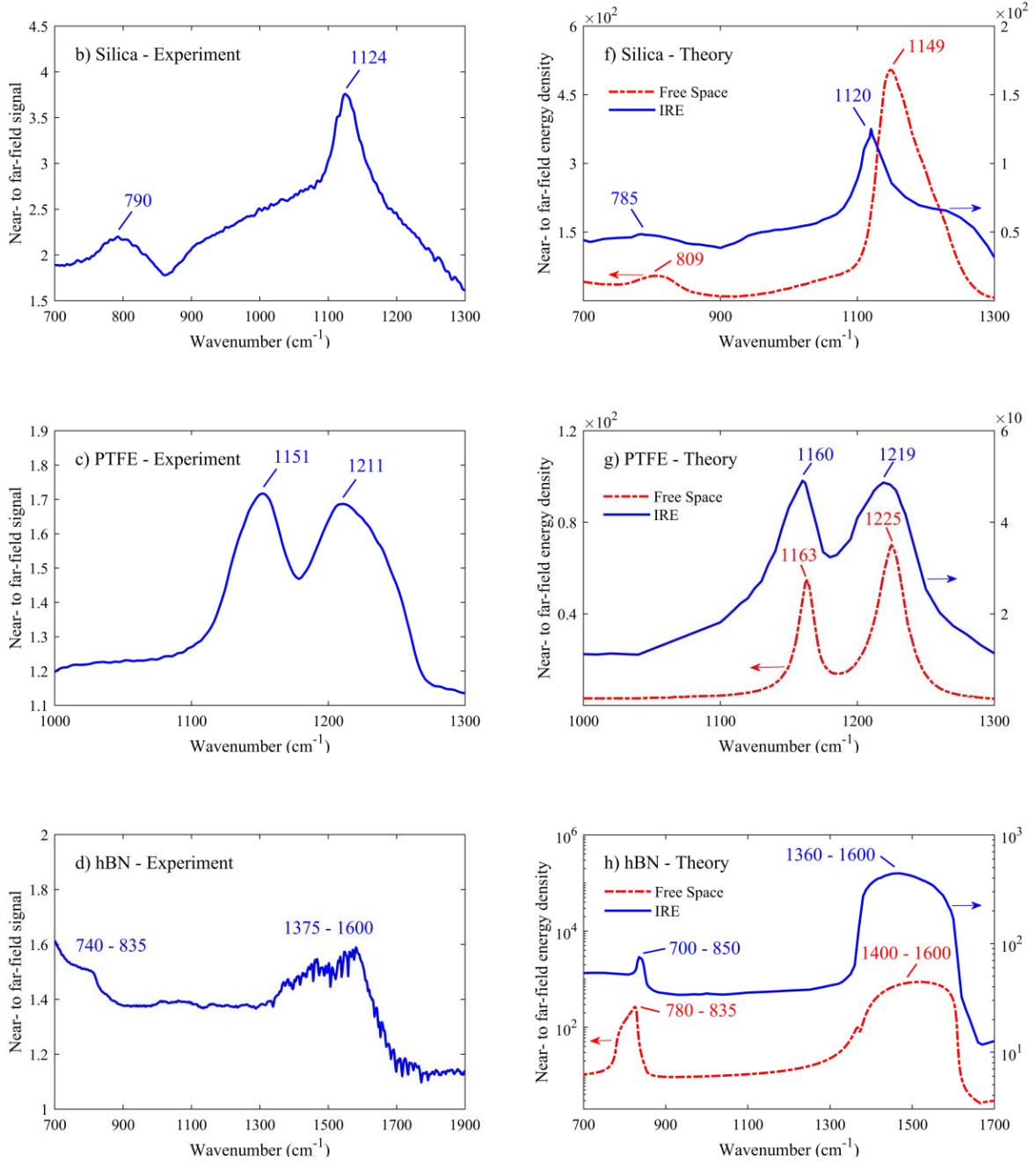

Figure 2 – The near-field thermal emission spectra normalized by the far-field value for quartz, silica, PTFE, and hBN. Panels (a) to (d) show the measured spectra, while Panels (e) to (h) display theoretically predicted spectra. In Panels (e) to (h), red dashed lines show the energy density at distance $d$ in the free space, while the solid blue lines show the energy density at distance $\Delta$ in the IRE due to the modes with $k_\rho$ between $\sin(20.4°)n_I k_0$ and $\sin(69.6°)n_I k_0$.

The measured near-field peaks are broadened and exhibit redshifts ranging from 8 cm$^{-1}$ to 42 cm$^{-1}$. Peak broadening and redshifts up to 105 cm$^{-1}$ are also observed in previously measured spectra [48-54]. The redshift and broadening in our measurements can be predicted by modeling energy density in the IRE due to thermal emission by the sample. The energy density at distance $\Delta$ in the IRE due to thermal emission by an anisotropic uniaxial medium separated by an air gap of size $d$ from the IRE (see the inset of Fig. 2e) is derived in Section V of Supplemental Materials. The energy density in the IRE is derived using the fluctuational electrodynamics and the DGFs for a one-dimensional layered media. It should be noted that we have verified that an air gap exists between the sample and the IRE (see Section VI of Supplemental Materials). The energy density of the modes exiting the IRE beveled ends, i.e., modes with $k_\rho$ between $\sin(20.4°)n_I k_0$ and $\sin(69.6°)n_I k_0$ (see Section I of Supplemental Materials), is derived as:

$$\langle u(\Delta,\omega) \rangle = \frac{\Theta(\omega,T_s)\omega}{2c_0^2\pi^2} \int_{\sin(69.6°)n_I k_0}^{\sin(20.4°)n_I k_0} k_\rho \int_0^{t_s} \left( k_I^2 \text{Trace}\left[ \bar{\bar{\mathbf{g}}}^E(k_\rho,\Delta,z',\omega) \cdot \text{Im}\left[\bar{\bar{\varepsilon}}_s\right] \cdot \bar{\bar{\mathbf{g}}}^{E\dagger}(k_\rho,\Delta,z',\omega) \right] \right. \\ \left. + \text{Trace}\left[ \bar{\bar{\mathbf{g}}}^H(k_\rho,\Delta,z',\omega) \cdot \text{Im}\left[\bar{\bar{\varepsilon}}_s\right] \cdot \bar{\bar{\mathbf{g}}}^{H\dagger}(k_\rho,\Delta,z',\omega) \right] \right) dz' dk_\rho \tag{1}$$

In Eq. 1, $u$ is the energy density, $\omega$ is the angular frequency, $\Theta$ is the mean energy of an electromagnetic state [57], $\langle\ \rangle$ represents ensemble average, $c_0$ is the speed of light in the free space, $T_s$ is the sample temperature, $t_s$ is the thickness of the sample, $k_I$ is the wavevector in the IRE, $z'$ is the vertical (to the surface) position of a thermally emitting source in the sample, superscript $\dagger$ indicates the Hermitian operator, $\bar{\bar{\varepsilon}}_s$ (= diag[$\varepsilon_{s,\perp}; \varepsilon_{s,\perp}; \varepsilon_{s,\parallel}$]) is the dielectric-function tensor for the sample, and $\bar{\bar{\mathbf{g}}}^E$ and $\bar{\bar{\mathbf{g}}}^H$ are the Weyl components of the electric ($E$) and magnetic ($H$) DGFs for an emitting anisotropic medium, respectively. The Weyl components of the DGFs are derived using the scattering matrix method [63] in Section V of Supplemental Materials.

The energy density in the middle of the IRE (where the signal is collected using the ATR accessory) is calculated in the near and far fields for a sample temperature of 160°C. The ratio of the near-field and far-field energy densities is shown in Figs. 2e to 2h for the four samples. The calculated spectra are in great agreement with the measurements showing that Eq. 1 can be used for predicting the redshift and broadening of the near-field peaks. These redshifts and broadening arise from the multiple reflections of the thermally emitted waves between the sample and the IRE as well as the fact that a portion of the near-field evanescent waves are captured in the measurements. However, all the SPhP, frustrated and hyperbolic modes are captured in the measured spectra.

In summary, we experimentally demonstrated narrow peaks of near-field thermal emission due to the excitation of SPhPs (for quartz and silica) and frustrated total-internal-reflection modes (for silica and PTFE), as well as the broadband thermal emission of hyperbolic modes in the near field (for hBN). We derived an analytical expression for the energy density inside the IRE due to thermal emission by an anisotropic medium that can be used for predicting the redshift and broadening of the measured near-field peaks. The presented spectroscopy technique can be used for measuring the spectra of near-field thermal emission by different materials.

**Acknowledgments**

The authors acknowledge support from the National Science Foundation under Grant No. CBET-1804360.

**Supplemental Materials for Article**

**The Measurement of Near-Field Thermal Emission Spectra using an Infrared Waveguide**

Saman Zare[*], Carl P. Tripp[**,†] and Sheila Edalatpour[*,†]

[*] *Department of Mechanical Engineering, University of Maine, Orono, ME 04469*

[**] *Department of Chemistry, University of Maine, Orono, ME 04469*

[†] *Frontier Institute for Research in Sensor Technologies, University of Maine, Orono, ME 04469*

**I. Parallel component of the wavevector for the thermally emitted waves exiting the IRE**

In this section, we determine the parallel component of the wavevector, $k_\rho$, for the thermally emitted waves that can exit the IRE beveled ends and reach the FTIR spectrometer. In the following sub-sections, we analyze the interaction of the waves having wavevectors $k_\rho < k_0$, $k_0 < k_\rho < n_I k_0$, and $k_\rho > n_I k_0$ with the IRE to determine if they can exit the IRE.

**A. Waves with $k_\rho < k_0$**

The emitted waves with $k_\rho < k_0$ are propagative in both the air gap and the IRE. The propagation of these waves toward the IRE beveled ends is schematically shown in Fig. 1a. These waves arrive at the air-IRE interface with an angle of $\theta_0$, which is between 0° and 90°, and partially transmit to the IRE (Fig. 1a). The transmitted waves make an angle of $\theta_1$ with the surface normal that can be found by considering the conservation of the parallel-component of the wavevector (Snell's law) as:

$$\theta_1 = \sin^{-1}\left(\frac{\sin\theta_0}{n_I}\right) \tag{S1}$$

In Eq. S1, $n_I$ is the refractive index of the IRE which is equal to 2.4 for ZnSe. Since $0° < \theta_0 < 90°$, $0° < \theta_1 < \theta_{cr}$ where $\theta_{cr} = \sin^{-1}(1/n_I)$ is the critical angle for the air-IRE interface ($\theta_{cr} = 24.6°$ for a ZnSe IRE). After multiple reflections at the air-IRE interface, the propagative waves hit the IRE beveled ends at an angle of $\theta_2 = 90° - (\alpha + \theta_1)$, where $\alpha$ is the bevel angle of the IRE. Considering that $0° < \theta_1 < 24.6°$ and $\alpha = 45°$ in our experiments, the lower and upper limits for $\theta_2$ are found as 20.4° and 45°, respectively. The waves with $\theta_2 \geq \theta_{cr} = 24.6°$ totally internally reflect at the beveled surface, while the waves with $20.4° < \theta_2 < 24.6°$ exit the IRE. The waves exiting the IRE have a $k_\rho$ in the range of $\sin(20.4°)\, n_I k_0 < k_\rho < k_0$.

**B. Waves with $k_0 < k_\rho < n_I k_0$**

The interaction of the waves having $k_\rho$ in the range of $k_0 < k_\rho < n_I k_0$ with the IRE is schematically shown in Fig. 1b. These waves are evanescent in the air but become propagative in the IRE (see Fig. 1b). The parallel component of the wavevector can be written as $k_\rho = n_I k_0 \sin \theta_1$. As such, the angle of propagation in the IRE is:

$$\theta_1 = \sin^{-1}\left(\frac{k_\rho}{n_I k_0}\right) \tag{S2}$$

Using Eq. 2S and considering waves with $k_0 < k_\rho < n_I k_0$, it is found that $\theta_{cr} = 24.6° < \theta_1 < 90°$. Since $\theta_1 > \theta_{cr}$, these waves experience multiple total internal reflections at the air-IRE interfaces until they reach the beveled surface at angle $\theta_2$. By a geometrical analysis, it can be shown that $\theta_2 = 90° - (\alpha + \theta_1)$ when $24.6° < \theta_1 < 45°$ and $\theta_2 = (\alpha + \theta_1) - 90°$ when $45° < \theta_1 < 90°$. Considering $\alpha = 45°$, the range of $\theta_2$ for the former and latter cases is found to be $0 < \theta_2 < 20.4°$ and $0 < \theta_2 < 45°$, respectively. Since $\theta_2 < \theta_{cr} = 24.6°$ in the former case, all these waves can exit the beveled surface. These waves have a $k_\rho$ in the range of $k_0 < k_\rho < \sin(45°) n_I k_0$. In the latter case, only the waves for which $0 < \theta_2 < \theta_{cr} = 24.6°$ can transmit to the air from the beveled surface. The parallel component of the wavevector for these waves varies in the range of $\sin(45°)n_I k_0 < k_\rho < \sin(69.6°)n_I k_0$. In total, the waves with $k_0 < k_\rho < \sin(69.6°)n_I k_0$ can exit the beveled side of the IRE and reach the FTIR spectrometer.

**C. Waves with $k_\rho > n_I k_0$**

The waves with $k_\rho > n_I k_0$ cannot propagate in the IRE as they are evanescent in both the air and the IRE. These waves do not exit from the IRE beveled surface.

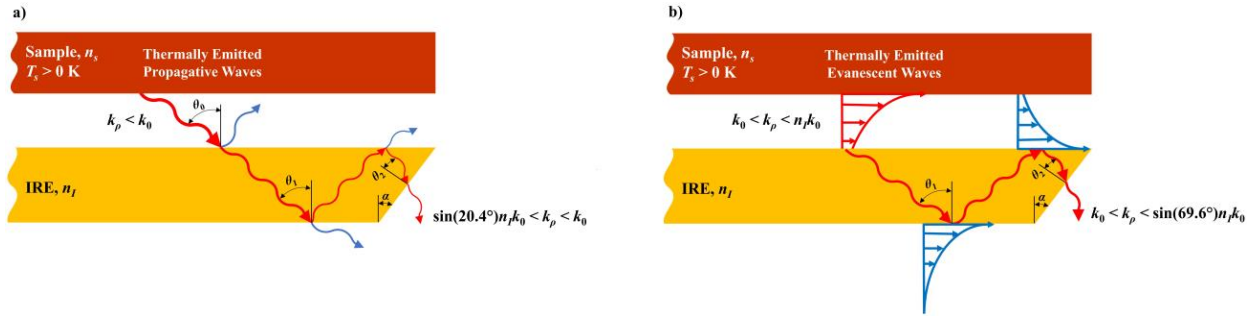

Figure 1 – The interaction of the thermally emitted waves with $k_\rho < k_0$ and $k_0 < k_\rho < n_I k_0$ with the IRE. (a) The waves with $k_\rho < k_0$ are propagative in both the air and the IRE. From these waves, those with $\sin(20.4°) n_I k_0 < k_\rho < k_0$ exit the IRE. (b) The waves with $k_0 < k_\rho < n_I k_0$ are evanescent in the air but propagative in the IRE. From these waves, those with $k_0 < k_\rho < \sin(69.6°) n_I k_0$ exit the IRE.

To conclude, a thermally emitted wave can exit the IRE and be collected by the FTIR if its parallel component of wavevector, $k_\rho$, is in the range of $\sin(20.4°)n_I k_0 < k_\rho < \sin(69.6°)n_I k_0$.

## II. The temperature of the sample

The voltage supplied to the heater is adjusted such that the sample reaches a temperature of 160°C in the far-field measurements. If the same voltage is supplied in the near-field measurements, the sample temperature drops below 160°C due to the conductive heat transfer with the IRE. As such, the supplied voltage needs to be increased for near-field measurements. The sample temperature cannot be measured in near-field experiments because the sample is in contact with the IRE. We increased the supplied voltage for near-field measurements until the ratio of the near-field and far-field spectra becomes flat at large wavenumbers where thermal emission is negligible due to the low energy of thermal oscillators. In spectroscopy, the spectral location of the peaks is of interest rather than the magnitude of thermal emission. Therefore, it is not required to have the exact same temperature for both near-field and far-field experiments. We experimentally verified that the spectral locations of the peaks are not affected by changing the supplied voltage in the near-field measurements.

**III. Compensation for background thermal emission, wavenumber-dependent responsivity of the photodetector, absorption by internal parts of the ATR accessory and the ambient, and modulation efficiency of the FTIR spectrometer**

The intensity of the measured signal, $I$, can be written as:

$$I(\omega) = \beta(\omega)\left[I^{s}(\omega) + I^{bk}(\omega)\right] \tag{S3}$$

where $\beta$ is a frequency-dependent factor that accounts for the absorption by the internal components of the ATR accessory and the ambient gases, the modulation efficiency of the FTIR spectrometer, and the responsivity of the photodetector, $I^s$ is the intensity of thermal emission by the sample, and $I^{bk}$ is the intensity of the background thermal emission. The factor $\beta$ varies with frequency, but it remains the same at a given frequency for all measurements.

To measure the ratio of near-field and far-field thermal emission by the sample, first the background thermal radiation is recorded. The background emission is measured by blocking thermal radiation by the sample from reaching the FTIR via placing a thick film of stainless steel (which is opaque in the infrared) at the exit of the ATR accessory. In this case, $I^s = 0$ and the measured signal, $I$, equals $\beta I^{bk}$ (see Eq. S3). Then, the block is removed, and the signal is recorded for the cases where the sample is in contact (near-field) and at 1-mm distance (far-field) from the IRE. Using Eq. S3, the intensity of the recorded signals can be written as:

$$I^{NF}(\omega) = \beta(\omega)\left[I^{s,NF}(\omega) + I^{bk}(\omega)\right] \tag{S4-a}$$

$$I^{FF}(\omega) = \beta(\omega)\left[I^{s,FF}(\omega) + I^{bk}(\omega)\right] \tag{S4-b}$$

where $I^{NF}$ shows the intensity of the signal for the near-field measurement, $I^{FF}$ shows the intensity of the signal for the far-field measurement, and $I^{s,NF}$ and $I^{s,FF}$ indicate the intensity of near-field and far-field thermal emission by the sample, respectively. The ratio $I^{s,NF}/I^{s,FF}$ is obtained by subtracting the background signal ($\beta I^{bk}$) from the near- and far-field signals ($I^{NF}$ and $I^{FF}$, respectively) and taking the ratio of the reduced signals, i.e.,

$$\frac{I^{s,NF}}{I^{s,FF}} = \frac{I^{NF}(\omega) - \beta(\omega)I^{bk}(\omega)}{I^{FF}(\omega) - \beta(\omega)I^{bk}(\omega)} \tag{S5}$$

### IV. Dielectric functions of quartz, silica, polytetrafluoroethylene, and hexagonal boron nitride

The dielectric functions of quartz [2], silica [2], polytetrafluoroethylene (PTFE) [3], and hexagonal boron nitride (hBN) [4] are obtained from the experimental data in literature. The dielectric function of hBN in the parallel and perpendicular directions (relative to the optical axis) is written using the Lorentz oscillator model as $\varepsilon_s = \varepsilon_\infty \left(\omega^2 - \omega_{LO}^2 + i\Gamma\omega\right) / \left(\omega^2 - \omega_{TO}^2 + i\Gamma\omega\right)$. The experimental oscillator parameters for parallel dielectric function are $\varepsilon_\infty = 2.95$, $\omega_{LO} = 830$ cm$^{-1}$, $\omega_{TO} = 780$ cm$^{-1}$, and $\Gamma = 4$ cm$^{-1}$, while for the perpendicular dielectric function they are $\varepsilon_\infty = 4.87$, $\omega_{LO}$ = 1610 cm$^{-1}$, $\omega_{TO}$ = 1370 cm$^{-1}$, and $\Gamma$ = 5 cm$^{-1}$ [4]. The dielectric functions of the samples are plotted versus the wavenumber in Fig. 2.

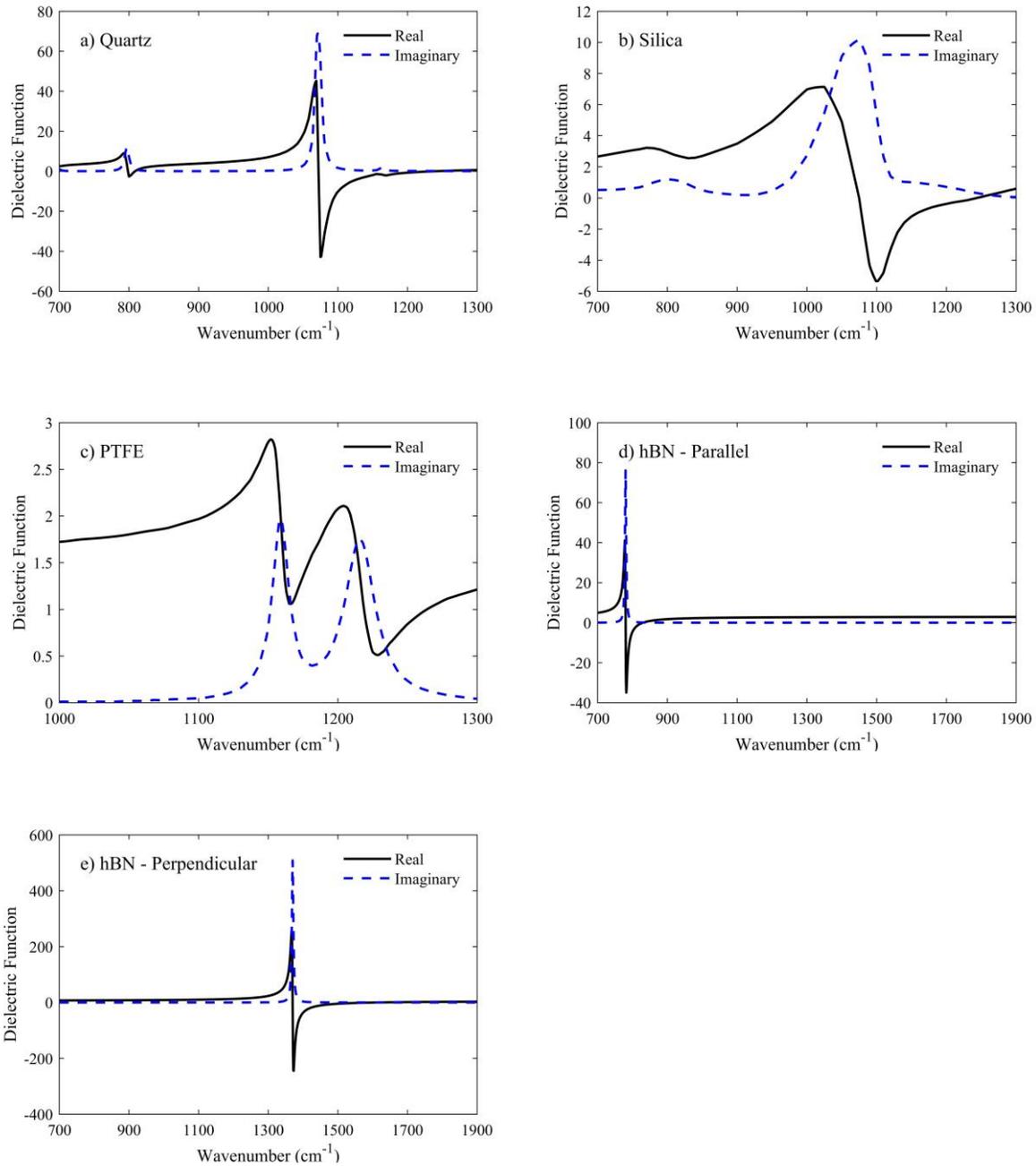

Figure 2 – The dielectric functions of (a) quartz, (b) silica, (c) PTFE, and (d) and (e) hBN.

## V. Energy density emitted by an anisotropic, uniaxial medium into the IRE

As discussed in Section I, the waves with a parallel component of wavevector, $k_\rho$, between $\sin(20.4°)\, n_I k_0$ and $\sin(69.6°) n_I k_0$ can exit the IRE and be captured by the FTIR spectrometer. The energy density of these waves in the IRE approximately models the measured spectrum. The energy density at distance $\Delta$ in the IRE can be obtained using the dyadic Green's functions for one-dimenisonal layered media and scattering matrix method. A schematic of the problem under consideration is shown in Fig. 3. An anisotropic uniaxial sample with thickness $t_s$ and temperature $T_s$ is emitting thermal radiation. The anisotropic dielectric response of the sample is described using a frequency-dependent diagonal tensor as $\overline{\overline{\varepsilon}}_s = \mathrm{diag}\,[\varepsilon_{s,\perp};\varepsilon_{s,\perp};\varepsilon_{s,\parallel}]$, where $\varepsilon_{s,\perp}$ and $\varepsilon_{s,\parallel}$ are the dielectric functions of the sample perpendicular and parallel to the optical axis ($z$-axis) of the sample. For isotropic samples (quartz, silica, and PTFE), $\varepsilon_{s,\parallel} = \varepsilon_{s,\perp} = \varepsilon_s$. An IRE with thickness $t_I$ and dielectric function $\varepsilon_I = n_I^2$ which is transparent in the wavenumber range of interest (Im[$\varepsilon_I$] $\approx 0$) is placed at distance $d$ from the sample. Thermal emission by the IRE is negligible compared to that of the sample as it is transparent. The sample, IRE and free space are labeled as $s$, $I$ and $v$, respectively. The objective is to find the energy density at distance $\Delta$ in the IRE.

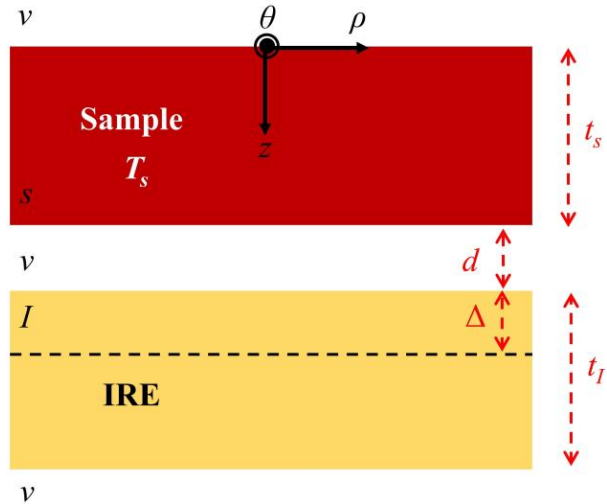

Figure 3 – A Schematic of the problem under consideration. Energy density at distance $\Delta$ in the IRE due to thermal emission by the heated sample is desired.

The time-averaged spectral density in the IRE is given as [5]:

$$\langle u(\mathbf{r},\omega)\rangle = \frac{1}{4}\varepsilon_I\varepsilon_0 \text{Trace}\langle \mathbf{E}(\mathbf{r},\omega)\otimes \mathbf{E}(\mathbf{r},\omega)\rangle + \frac{1}{4}\mu_0 \text{Trace}\langle \mathbf{H}(\mathbf{r},\omega)\otimes \mathbf{H}(\mathbf{r},\omega)\rangle \quad \text{(S6)}$$

where $u$ is the energy density, $\omega$ is the angular frequency, $\mathbf{r}$ is the position at which the energy density is desired, $\langle\;\rangle$ represents the ensemble average, $\otimes$ is the outer product, $\varepsilon_0$ and $\mu_0$ are the permittivity and permeability of the free space, respectively, and $\mathbf{E}$ and $\mathbf{H}$ are the electric and magnetic fields emitted by the heated sample, respectively. The electric and magnetic fields can be obtained using the electric and magnetic dyadic Green's functions, $\bar{\bar{\mathbf{G}}}^E$ and $\bar{\bar{\mathbf{G}}}^H$ and the thermally stochastic current $\mathbf{J}^{fl}$ as [6]:

$$\mathbf{E}(\mathbf{r},\omega) = i\omega\mu_0 \int_{V_s} \bar{\bar{\mathbf{G}}}^E(\mathbf{r},\mathbf{r}',\omega)\cdot \mathbf{J}^{fl}(\mathbf{r}',\omega)dV' \quad \text{(S7-a)}$$

$$\mathbf{H}(\mathbf{r},\omega) = \int_{V_s} \bar{\bar{\mathbf{G}}}^H(\mathbf{r},\mathbf{r}',\omega)\cdot \mathbf{J}^{fl}(\mathbf{r}',\omega)dV' \quad \text{(S7-b)}$$

where the integral is performed over the volume of the sample, $V_s$, where the fluctuating current is non-zero. The fluctuating current is given by the fluctuation-dissipation theorem as [7,8]:

$$\langle \mathbf{J}^{fl}(\mathbf{r}',\omega)\otimes \mathbf{J}^{fl}(\mathbf{r}'',\omega')\rangle = \frac{4}{\pi}\omega\varepsilon_0\,\text{Im}\left[\bar{\bar{\varepsilon}}_s\right]\Theta(\omega,T_s)\delta(\mathbf{r}'-\mathbf{r}'')\delta(\omega-\omega') \quad \text{(S8)}$$

where $\Theta$ is the mean energy of an electromagnetic state [8]. By substituting Eqs. S7 and S8 into Eq. S6, the energy density can be written as:

$$\langle u(\mathbf{r},\omega)\rangle = \frac{k_0^2}{\pi\omega}\Theta(\omega,T)\int_{V_s}\text{Trace}\Big[k_0^2\overline{\overline{\mathbf{G}}}^E(\mathbf{r},\mathbf{r}')\cdot\text{Im}\big[\overline{\overline{\varepsilon}}_s\big]\cdot\overline{\overline{\mathbf{G}}}^{E\dagger}(\mathbf{r},\mathbf{r}')$$
$$+\overline{\overline{\mathbf{G}}}^H(\mathbf{r},\mathbf{r}')\cdot\text{Im}\big[\overline{\overline{\varepsilon}}_s\big]\cdot\overline{\overline{\mathbf{G}}}^{H\dagger}(\mathbf{r},\mathbf{r}')\Big]dV' \quad (S9)$$

where superscript † indicates the Hermitian operator. The dyadic Green's functions can be expressed as an integral of plane waves using the Weyl representation as [9]:

$$\overline{\overline{\mathbf{G}}}^{E(H)}(\mathbf{r},\mathbf{r}') = \int_{-\infty}^{\infty}\overline{\overline{\mathbf{g}}}^{E(H)}(k_\rho,z,z',\omega)e^{ik_\rho(\mathbf{R}-\mathbf{R}')}d\mathbf{k}_\rho \quad (S10)$$

where $\overline{\overline{\mathbf{g}}}^{E(H)}$ is the Weyl component of the electric (magnetic) dyadic Green's function, $\mathbf{k}_\rho$ is the parallel (to the surface) component of the wavevector, and $\mathbf{R}$ ($z$) and $\mathbf{R}'$ ($z'$) are the parallel (perpendicular) components of the position vectors of the observation and source points, respectively. Substituting Eq. S10 into Eq. S9 and exploiting the azimuthal symmetry of the geometry, the energy density at distance $\Delta$ in the IRE due to the waves with $k_\rho$ between $\sin(20.4°)n_I k_0$ and $\sin(69.6°)n_I k_0$ is written as:

$$\langle u(\Delta,\omega)\rangle = \frac{\Theta(\omega,T_s)\omega}{2c_0^2\pi^2}\int_{\sin(20.4°)n_I k_0}^{\sin(69.6°)n_I k_0}k_\rho\int_0^{t_s}\Big(k_I^2\text{Trace}\Big[\overline{\overline{\mathbf{g}}}^E(k_\rho,\Delta,z',\omega)\cdot\text{Im}\big[\overline{\overline{\varepsilon}}_s\big]\cdot\overline{\overline{\mathbf{g}}}^{E\dagger}(k_\rho,\Delta,z',\omega)$$
$$+\overline{\overline{\mathbf{g}}}^H(k_\rho,\Delta,z',\omega)\cdot\text{Im}\big[\overline{\overline{\varepsilon}}_s\big]\cdot\overline{\overline{\mathbf{g}}}^{H\dagger}(k_\rho,\Delta,z',\omega)\Big]\Big)dz'dk_\rho \quad (S11)$$

The Weyl components of the electric and magnetic Green's functions can be written using the Sipe unit vectors as [10]:

$$\bar{\bar{\mathbf{g}}}^E\left(k_\rho, \Delta, z', \omega\right) = \frac{i}{2k_{z,s}^{TE}}\left(A_I^{TE} e^{ik_{z,I}\Delta - ik_{z,s}^{TE} z'} + B_I^{TE} e^{-ik_{z,I}\Delta - ik_{z,s}^{TE} z'} + C_I^{TE} e^{ik_{z,I}\Delta + ik_{z,s}^{TE} z'} + D_I^{TE} e^{-ik_{z,I}\Delta + ik_{z,s}^{TE} z'}\right)\hat{\mathbf{s}}\hat{\mathbf{s}}$$

$$+ \frac{i}{2k_{z,s}^{TM}}\left(A_I^{TM} e^{ik_{z,I}\Delta - ik_{z,s}^{TM} z'}\hat{\mathbf{p}}_I^+\hat{\mathbf{p}}_s^+ + B_I^{TM} e^{-ik_{z,I}\Delta - ik_{z,s}^{TM} z'}\hat{\mathbf{p}}_I^-\hat{\mathbf{p}}_s^+ + C_I^{TM} e^{ik_{z,I}\Delta + ik_{z,s}^{TM} z'}\hat{\mathbf{p}}_I^+\hat{\mathbf{p}}_s^- + D_I^{TM} e^{-ik_{z,I}\Delta + ik_{z,s}^{TM} z'}\hat{\mathbf{p}}_I^-\hat{\mathbf{p}}_s^-\right)$$

(S12-a)

$$\bar{\bar{\mathbf{g}}}^H\left(k_\rho, \Delta, z', \omega\right) = \frac{k_I}{2k_{z,s}^{TE}}\left(A_I^{TE} e^{ik_{z,I}\Delta - ik_{z,s}^{TE} z'}\hat{\mathbf{p}}_I^+\hat{\mathbf{s}} + B_I^{TE} e^{-ik_{z,I}\Delta - ik_{z,s}^{TE} z'}\hat{\mathbf{p}}_I^-\hat{\mathbf{s}} + C_I^{TE} e^{ik_{z,I}\Delta + ik_{z,s}^{TE} z'}\hat{\mathbf{p}}_I^+\hat{\mathbf{s}} + D_I^{TE} e^{-ik_{z,I}\Delta + ik_{z,s}^{TE} z'}\hat{\mathbf{p}}_I^-\hat{\mathbf{s}}\right)$$

$$- \frac{k_I}{2k_{z,s}^{TM}}\left(A_I^{TM} e^{ik_{z,I}\Delta - ik_{z,s}^{TM} z'}\hat{\mathbf{s}}\hat{\mathbf{p}}_s^+ + B_I^{TM} e^{-ik_{z,I}\Delta - ik_{z,s}^{TM} z'}\hat{\mathbf{s}}\hat{\mathbf{p}}_s^+ + C_I^{TM} e^{ik_{z,I}\Delta + ik_{z,s}^{TM} z'}\hat{\mathbf{s}}\hat{\mathbf{p}}_s^- + D_I^{TM} e^{-ik_{z,I}\Delta + ik_{z,s}^{TM} z'}\hat{\mathbf{s}}\hat{\mathbf{p}}_s^-\right)$$

(S12-b)

In Eq. S12, superscript *TE* (*TM*) refers to the transverse electric (magnetic) polarization, and $\hat{\mathbf{s}}$ and $\hat{\mathbf{p}}_i^\pm$ are the Sipe unit vectors for TE- and TM-polarizations inside layer *i*, respectively. The Sipe unit vectors in the sample and the IRE are given by [10,11]:

$$\hat{\mathbf{s}} = -\hat{\boldsymbol{\theta}} \tag{S13-a}$$

$$\hat{\mathbf{p}}_s^\pm = \frac{1}{k_s^{TE}}\left(\mp k_{z,s}^{TM}\hat{\boldsymbol{\rho}} + \frac{\varepsilon_{s,\perp}}{\varepsilon_{s,\parallel}}k_\rho\hat{\mathbf{z}}\right) \tag{S13-b}$$

$$\hat{\mathbf{p}}_I^\pm = \frac{1}{k_I}\left(\mp k_{z,I}\hat{\boldsymbol{\rho}} + k_\rho\hat{\mathbf{z}}\right) \tag{S13-c}$$

where $k_s^{TE} = \sqrt{\varepsilon_{s,\perp}}k_0$ is the magnitude of the TE-polarized wavevector in the sample and $k_{z,s}^{TM} = \sqrt{\varepsilon_{s,\perp}k_0^2 - \frac{\varepsilon_{s,\perp}}{\varepsilon_{s,\parallel}}k_\rho^2}$ is the *z*-component of the TM-polarized wavevector in the sample.

The coefficients $A_I^\gamma$ ($B_I^\gamma$) and $C_I^\gamma$ ($D_I^\gamma$) are the amplitude of the $\gamma$-polarized ($\gamma$ = *TE* or *TM*) waves in the IRE traveling toward the positive (negative) direction of the *z*-axis due to thermal sources emitting in the positive and negative directions of the *z*-axis, respectively [10]. The coefficients $A_I^\gamma$, $B_I^\gamma$, $C_I^\gamma$ and $D_I^\gamma$ can be found using the scattering matrix method as [10]:

$$A_I^\gamma = \frac{t_{sv}^\gamma t_{vI}^\gamma e^{ik_{z,s}^\gamma t_s} e^{ik_{z,v}^\gamma d}}{\left(1 + r_{vs}^\gamma r_{sv}^\gamma e^{2ik_{z,s}^\gamma t_s}\right)\left(1 + r_{vI}^\gamma r_{Iv}^\gamma e^{2ik_{z,I}^\gamma t_I}\right)\left(1 - R_s^\gamma R_I^\gamma e^{2ik_{z,v}^\gamma d}\right)} \quad \text{(S14-a)}$$

$$B_I^\gamma = r_{Iv}^\gamma e^{2ik_{z,I}^\gamma t_I} A_I^\gamma \quad \text{(S14-b)}$$

$$C_I^\gamma = -r_{vs}^\gamma A_I^\gamma \quad \text{(S14-c)}$$

$$D_I^\gamma = r_{Iv}^\gamma e^{2ik_{z,I}^\gamma t_I} C_I^\gamma \quad \text{(S14-d)}$$

where and $r_{ij}^\gamma$ and $t_{ij}^\gamma$ are the Fresnel reflection and transmission coefficients at the interface of layers *i* and *j* for *γ*-polarization, respectively, The Fresnel coefficients are given by [11]:

$$r_{ij}^{TE} = \frac{k_{z,i}^{TE} - k_{z,j}^{TE}}{k_{z,i}^{TE} + k_{z,j}^{TE}} \quad \text{(S15-a)}$$

$$r_{ij}^{TM} = \frac{\varepsilon_{j,\perp} k_{z,i}^{TM} - \varepsilon_{i,\perp} k_{z,j}^{TM}}{\varepsilon_{j,\perp} k_{z,i}^{TM} + \varepsilon_{i,\perp} k_{z,j}^{TM}} \quad \text{(S15-b)}$$

$$t_{ij}^{TE} = \frac{2k_{z,i}^{TE}}{k_{z,i}^{TE} + k_{z,j}^{TE}} \quad \text{(S15-c)}$$

$$t_{ij}^{TM} = \frac{2\varepsilon_{j,\perp} k_{z,i}^{TM}}{\varepsilon_{j,\perp} k_{z,i}^{TM} + \varepsilon_{i,\perp} k_{z,j}^{TM}} \sqrt{\frac{\varepsilon_{i,\perp}}{\varepsilon_{j,\perp}}} \quad \text{(S15-d)}$$

In Eq. S15, $k_{z,i}^{TE}$ is the z-component of the TE-polarized wave in the IRE and is given by $k_{z,i}^{TE} = \sqrt{\varepsilon_{i,\perp} k_0^2 - k_\rho^2}$. Note that for an isotropic sample $k_{z,i}^{TE} = k_{z,i}^{TM}$. In Eq. S14, $R_j^\gamma$ represents the reflection coefficient of layer *j* in the free space for polarization state $\gamma$ and is found as [12]:

$$R_j^\gamma = \frac{r_{vj}^\gamma + r_{jv}^\gamma e^{2ik_{z,j}^\gamma t_j}}{1 + r_{vj}^\gamma r_{jv}^\gamma e^{2ik_{z,j}^\gamma t_j}} \tag{S16}$$

## VI. Verification of the existence of an air gap between the sample and the IRE in the experiments

In this section, we verify that an air gap exists between the IRE and the sample in our experiments. We compute the energy density in the middle of the IRE ($\Delta = 1$ mm) due to the waves with $k_\rho$ between $\sin(20.4°) n_I k_0$ and $\sin(69.6°) n_I k_0$ using Eq. S11 for two cases. In the first case, a gap of $d = 200$ nm is assumed between the sample and the IRE, while no gap ($d = 0$) is assumed in the second case. The quartz sample at a temperature of $T_s = 160°C$ is considered. The computed spectra with $d = 200$ nm and 0 are compared with the measured spectrum in Fig. 4. While the measured spectrum agrees well with the computed energy density at $d = 200$ nm, it is significantly different from that calculated for $d = 0$. This verifies that an air gap exists between the sample and the IRE when they are in contact.

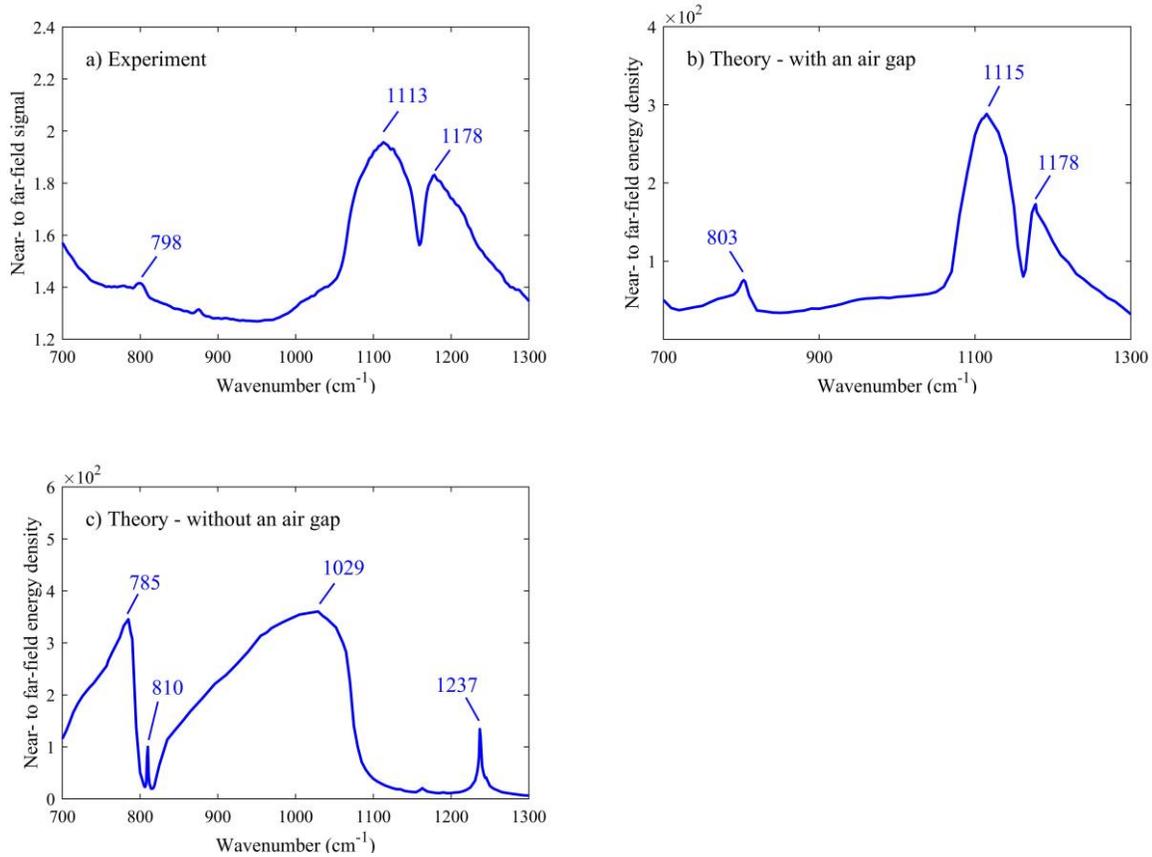

Figure 4 – The ratio of near-field and far-field thermal spectra for quartz at 160°C. Panel (a) shows the measured spectrum, while the theoretical spectra at $d$ = 200 nm and 0 are presented in Panels (b) and (c), respectively.